\documentclass[aps,prl,10pt,twocolumn,superscriptaddress]{revtex4-1}
\usepackage{graphicx}
\usepackage{epsfig}
\usepackage{amsmath}
\usepackage{amssymb}
\usepackage{amsfonts}
\usepackage{mathrsfs}
\usepackage{theorem}
\usepackage{bm}
\usepackage{url}
\usepackage{enumerate}
\usepackage[T1]{fontenc}
\usepackage{csquotes}
\usepackage{natbib}
\usepackage{float}
\usepackage{appendix}
\usepackage{booktabs}
\usepackage{soul}
\usepackage{color}
\usepackage[colorlinks=true,citecolor=blue,linkcolor=blue]{hyperref}
\MakeOuterQuote{"}

\begin{document}


\title{Advances in quantum dense coding}


\author{Yu Guo}
\affiliation{CAS Key Laboratory of Quantum Information, University of Science and Technology of China, Hefei, 230026, People's Republic of China.}
\affiliation{CAS Center For Excellence in Quantum Information and Quantum Physics, University of Science and Technology of China, Hefei 230026, P.R. China.}

\author{Bi-Heng Liu}
\email{bhliu@ustc.edu.cn}
\affiliation{CAS Key Laboratory of Quantum Information, University of Science and Technology of China, Hefei, 230026, People's Republic of China.}
\affiliation{CAS Center For Excellence in Quantum Information and Quantum Physics, University of Science and Technology of China, Hefei 230026, P.R. China.}

\author{Chuan-Feng Li}
\affiliation{CAS Key Laboratory of Quantum Information, University of Science and Technology of China, Hefei, 230026, People's Republic of China.}
\affiliation{CAS Center For Excellence in Quantum Information and Quantum Physics, University of Science and Technology of China, Hefei 230026, P.R. China.}

\author{Guang-Can Guo}
\affiliation{CAS Key Laboratory of Quantum Information, University of Science and Technology of China, Hefei, 230026, People's Republic of China.}
\affiliation{CAS Center For Excellence in Quantum Information and Quantum Physics, University of Science and Technology of China, Hefei 230026, P.R. China.}

\date{\today}

\begin{abstract}
Quantum dense coding is one of the most important protocols in quantum communication. It derives from the idea of using quantum resources to boost the communication capacity and now serves as a key primitive across a variety of quantum information protocols. Here, we focus on the basic theoretical ideas behind quantum dense coding, discussing its development history from discrete and continuous variables to quantum networks, then to its variant protocols and applications in quantum secure communication. With this basic background in hand, we then review the main experimental achievements, from photonic qubits and qudits to optical modes, nuclear magnetic resonance, and atomic systems. Besides the state of the art, we finally discuss potential future steps.
\end{abstract}

\maketitle
\section*{1. Introduction}
Human beings tried in various ways to exchange information and ideas with each others, even before the creation of languages. Vast information carriers, ranging from number of macrams, acoustic waves, characters and so on, have been used to convey signals from one site to another. One of the present directions in quantum science is promoting the communication efficiency, i.e. conveying more information with less carriers, which should now be treated as quantum particles. Such tasks are possible with the help of quantum resource.

In 1992, a seminal paper~\cite{1996Bennett}, described a quantum communication scheme, dubbed quantum dense coding or superdense coding later, that aimed to send more than one bit of classical information per every two-state particle transmitted. In this scheme, two classical bits (cbit) are allowed to be encoded reliably in one quantum bit (qubit) from an initial Einstein-Podolsky-Rosen (EPR) pair (ebit)~\cite{1935epr}, and therefore the classical capacity of a quantum channel is doubled if only the particle is transmitted and all the messages can be retrieved. The process can be summarized as:
\begin{equation}\label{densecoding}
1 qubit + 1 ebit = 2 cbits
\end{equation}
The most important resource required in this process is quantum entanglement~\cite{cventanglement,dventanglement}. In fact, quantum dense coding can be seen as a protocol that clearly demonstrates the character of quantum entanglement as a resource-- without its presence, such an enhancement of classical capacity would be impossible within the laws of quantum mechanics.

Quantum dense coding plays an important role in quantum information science~\cite{qinformation} in at least two aspects. First, it provides a paradigm of quantum communication, which, as mentioned above, provides a higher capacity than its counterpart in classical domain. Quantum dense coding has also been developed in a quantum network~\cite{2004Brubeta,2001Hao,2008Srerraman} and even has transcended classical message communication~\cite{2004Aram,06Hayden}. Second, it pours into the new vigor for other kinds of quantum communication. Quantum key distribution (QKD)~\cite{2002Long,2010Xihan} and also quantum secure direct communication (QSDC)~\cite{2002Beige,2003FuGuo} based on dense coding have been reported and have their advantages over pervious versions.

Recently, quantum dense coding has been realized in the labs with variety of information carriers from different quantum systems, including photonic qubits~\cite{96innsbruck}, optical modes~\cite{2002cvexperiment}, nuclear magnetic resonance (NMR)~\cite{2000nmr} and also atomic systems~\cite{2004atom,2017weizhang}. For photonic quantum dense coding, great efforts have been made to realize complete message retrieve~\cite{06experiment,08experiment,2017nankai,17experiment}. Another outstanding achievement has been made in terms of high-dimensional dense coding~\cite{2018experiment}, making it possible to realize even higher capacities. Dense coding in a quantum network has also been realized with optical~\cite{2003cvcontrol} modes and NMR~\cite{2004nmr}.

Here we provide an overview of quantum dense coding, including the basic theoretical ideas, its applications in quantum secure communication and, in particular, experimental progress in this field. For the experimental achievements, we mainly focus on dense coding protocols based on photonic and optical systems, for the excellent properties make them promising candidates for quantum communication. NMR and atomic systems based dense coding is also introduced.

\section*{2. Basic of quantum dense coding}
\begin{figure}[tbph]
	\includegraphics [width=0.95\columnwidth]{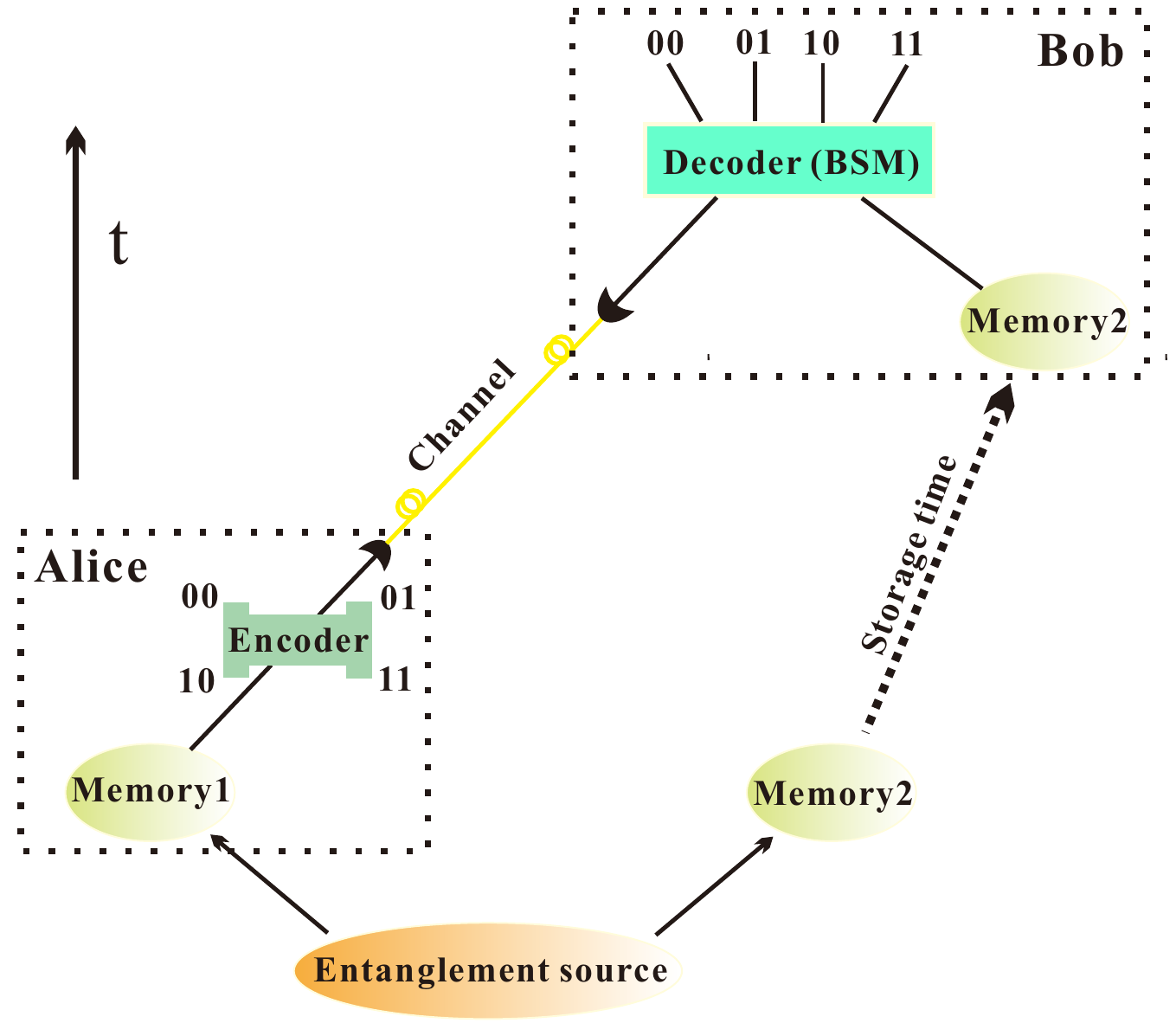}
	\caption{Scheme of quantum dense coding. Alice transmits a message to Bob with the help of a per-shared entangled state between them, which may be stored in their local quantum memories. Alice performs state transformations to encode her message and sends her quantum particle to Bob through a quantum channel which may not be noise-free. Bob decodes the message by performing Bell state measurements on his own particle and the particle from Alice. In qubit case, at most 2 bits can be transmitted with every qubit been sent. Variations based on this scheme may occur depending on the actual technologies adopted.}
	\label{fig:scheme}
\end{figure}

We start by giving a brief introduction of quantum dense coding protocol as shown in Fig.~\ref{fig:scheme} in several scenarios, including discrete variables (including qubit and qudit systems), continuous variables, and also multi-parties.

\bigskip
\noindent\textbf{2.1 Quantum dense coding of qubits}\\
Dense coding was originally raised on a two-level bipartite quantum systems~\cite{1996Bennett}. In this protocol, two remote parties, called Alice and Bob, want to communicate with each other by sending their pre-shared EPR pairs, such as the singlet state of two spin-$\frac{1}{2}$ particles,i.e. $|\Psi^-\rangle=\frac{1}{2}(|\uparrow\downarrow\rangle-|\downarrow\uparrow\rangle)$. When Alice want to convey information to Bob, She encodes her messages by performing one out of four possible unitary transformations from the Pauli set $\{\emph{I},\sigma_x,\sigma_y,\sigma_z\}$ on her own particle alone. Alice's operation on her particle may transform the shared singlet state into four Bell states, called signal states; specifically, $\sigma_x$ changes the state to $|\Phi^-\rangle=\frac{1}{2}(|\downarrow\downarrow\rangle-|\uparrow\uparrow\rangle)$, $\sigma_y$ changes the state to $|\Phi^+\rangle=\frac{1}{2}(|\downarrow\downarrow\rangle+|\uparrow\uparrow\rangle)$, $\sigma_z$ changes the state to $|\Psi^+\rangle=\frac{1}{2}(|\uparrow\downarrow\rangle+|\downarrow\uparrow\rangle)$, and identity operation makes no changes. The signal states with a priori probability compose the signal ensemble. After the encoding process, Alice then sends her particle to Bob through a quantum channel. Now possessing both particles, Bob decodes the messages by discrimination the Bell state of the pair, namely Bell state measurement (BSM). Since Alice's four kinds of unitary transformations result in four orthogonal Bell states, Bob can determine four distinguishable messages, i.e. 2 bits of classical information, based on his measurement outcomes. As a result, in the dense coding protocol, 2 bits of information is transmitted between two separated parties at a cost of transmitting a qubit carrier and consuming a pre-distributed ebit.

The major feature of encoding 2 bits of information by manipulating only one of the two particles is made possible by encode the message in superpositions of classical bipartite state combinations ($|\uparrow\downarrow\rangle$, $|\downarrow\uparrow\rangle$, $|\downarrow\downarrow\rangle$ and $|\uparrow\uparrow\rangle$). For these 4 classical combinations, we need manipulate both particles to encode 2 bits of message. In a communication protocol, the optimal amount of information that can be conveyed, i.e. channel capacity, is defined to be the maximum of the Holevo quantity~\cite{Kholevo1973} with respect to the signal ensemble, that is
\begin{equation}\label{capacity}
\emph{C}=\max\limits_{\{p_i,\rho_i\}}\chi=\max\limits_{\{p_i,\rho_i\}}[\emph{S}(\bar{\rho})-\sum_{i=0}^{i_{max}}p_i\emph{S}(\rho_i)]
\end{equation}
where $\emph{S}(\rho)=-\emph{Tr}(\rho \log{2}{\rho})$ denotes the von Neumann entropy and $\bar{\rho}=\sum_{i=0}^{i_{max}}p_i\rho_i$ is the average density matrix of the signal ensemble. The Holevo quantity $\chi$ can be also rewritten as in a quantum relative entropy~\cite{rmp2002,prl1997,pra1998} formula as $\chi=\sum_{i=0}^{i_{max}}p_i\emph{S}(\rho_i\parallel\bar{\rho})$, where $\emph{S}(\rho\parallel\sigma)$ is the quantum relative entropy of $\rho$ with respect to $\sigma$.

In ideal case, the capacity of a qubit dense coding protocol is 2 as discussed above. In practice, however, noise may affect the entanglement distribution and also the transmission of the signal states, resulting to a practical capacity that can never reach the theory limit of 2. These scenarios have been named as one-sided or two-sided noisy dense coding~\cite{2010Shadman}. In these case, it will need find the optimal signal ensemble to maximize the capacity $\emph{C}$. When the sender and the receiver share a mixed state entanglement, maximum $\chi$ can be yielded when the signal states are generated by mutually orthogonal unitary transformations with equal probabilities~\cite{2001Tohya}. With general a priori signal probabilities, a calculable limit on the capacity $\emph{C}$ can be imposed with the help of several entanglement measure~\cite{2001Tohya,2000Bose}. Mathematically, a noisy quantum channel can be described as a linear map $\Lambda$ acting on the quantum state. With such noisy channel, the dense coding capacity $\emph{C}$ can be obtained by replacing all quantum states in Eq.~(\ref{capacity}) with map $\Lambda$ on them~\cite{2010Shadman,2011zShadman}. Dense coding capacity $\emph{C}$ of the two-sided scenario has been derived for some special noise channels where the von Neumann entropy fulfills specific conditions~\cite{2010Shadman}. Quantum dense coding dealing with noisy channels has been reviewed~\cite{2013Shadman}. In a more general multi-round transmission scenario, adaptive strategies may be allowed to update the distributed and received states based on the result of previous rounds. In this case, Holevo quantity should be further optimized on the adaptive operations before the dense coding capacity is obtained~\cite{2019riccardo}.

For a valid dense coding protocol, the calculated dense coding capacity should exceed 1 which could be obtained in a perfect experiment (without consuming entanglement resource) where a quantum qubit is used to transmit classical information. As an example, bipartite bound entangled states cannot be used to construct a valid dense coding protocol~\cite{2004Brubeta,2001Horodecki}.

We next mention a variant of quantum dense coding, called probabilistic dense coding~\cite{1996Hausladen,2000Hao,2005Pati}. This protocol deals with the case when the sender and the receiver share non-maximally entangled pure state, which can be generally written as $\phi_{NME}=\alpha|00\rangle+\beta|11\rangle$ ($|\alpha|^2+|\beta|^2=1$, and suppose $|\alpha|<|\beta|$ here) under the Schmidt basis. There are mainly two ideas to realize probabilistic dense coding. One is by convert the non-maximally entangled states to a maximally entangled state (with probability $2|\beta|^2$)or a product state (with probability $1-2|\beta|^2$) with the help of an auxiliary qubit~\cite{2000Hao}. The result dense coding capacity $\emph{C}$ is $1+2|\beta|^2$. The other is by distinguishing four nonorthogonal signal states (also with probability $2|\beta|^2$) through the receiver's positive operator valued measurements. In this case, the receiver can extract 2 bits of classical information with a success probability given by $2|\beta|^2$.

Quantum dense coding may involve higher-dimensional quantum systems beyond qubits. A high-dimensional dense coding scheme~\cite{2002Liu} follows the lines of standard qubit dense coding, with the shared state replaced by a high-dimensional maximally entangled state and the Pauli operators replaced by high-dimensional Pauli operators~\cite{2001Werner}. Since there exists $d^2$ signal states, the receiver can recover $2\log_2 d$ bits of message after he implements complete d-dimensional Bell state measurements. More generally, a $m\times n$ dimensional system can be used to implement a dense coding protocol with theoretical capacity $\emph{C}=\log_2 (mn)$~\cite{2004Brubeta,2004Yan}.

\bigskip
\noindent\textbf{2.2 Quantum dense coding of continuous variables}\\
Quantum dense coding can be extended to continuous variable quantum systems~\cite{cvr2003,rmp2005,rmp2012}, which have an infinite dimensional Hilbert space. These systems are typically realized by optical modes, whose electromagnetic field can be described by position- and momentum-like quadrature operators, i.e. quadrature-phase amplitudes $(\hat{x}, \hat{p})$. In such continuous variable system, quantum states are drawn from the phase space for a single mode of the field and entangled resource is obtained from two beams (EPR beams) whose canonically conjugate variables $(\hat{x}, \hat{p})$ exist certain quantum correlations. Classical signal can be defined as $\alpha=\langle\hat{x}\rangle+i\langle\hat{p}\rangle$ and can then be directly associated with the quantum state $\hat{\rho}_{\alpha}$ of an optical mode.

In a typical continuous variable quantum dense coding protocol~\cite{1999Ban,2000Braunstein,2000Zhang}, Alice and Bob share EPR beams typically approximated by the two-mode squeezed state. Alice encodes her classical message $\alpha_{in}$ by implementing a phase-space offset by way of the displacement operator $\hat{D}(\alpha_{in})$ applied to her beam. The beam is then transmitted along a quantum channel to Bob, who then decodes the message by applying a continuous variable Bell measurement on his two modes. Mainly two methods for continuous variable Bell measurement were developed: one by mixing the two beams on a balanced beamsplitter and measuring the output ports with two homodyne detectors~\cite{1999Ban,2000Braunstein}; another by adding a phase shift of $\pi/2$ before combining the beams on a balanced beamsplitter and the signal can be detected with two photodetectors and two rf splitters~\cite{2000Zhang}. The obtained message $\alpha_{out}$ equals $\alpha_{in}$ in the limit $\bar{n}\rightarrow\infty$, where $\bar{n}$ is the number of photons associated with modulation and squeezing in the signal channel. This implies the classical message of the sender can be perfectly recovered by the receiver.

The dense coding capacity of the two-mode squeezed state protocol is given by $\emph{C}=\ln(1+\bar{n}+\bar{n^2})$~\cite{2000Braunstein}, which always beats single-mode coherent-state communication ($C_{coh}=\ln{(1+\bar{n})}$)~\cite{coh} and surpasses single-mode squeezed-state communication ($C_{sq}=\ln{(1+2\bar{n})}$)~\cite{sq} for $\bar{n}>1$ and also beats the optimal single channel communication ($C_{opt}=(1+\bar{n})\ln{(1+2\bar{n})}-\bar{n}\ln{\bar{n}}$)~\cite{opt1} for $\bar{n}\simeq1.884$. In the ideal case where perfect EPR beams are shared, the dense coding scheme allows twice as much messages to be encoded within a given state.
The continuous variable dense coding protocols allow unconditional signal transmission with high efficiency, in contrast to the conditional transmission with extremely low efficiency, caused by weak parametric down conversion, that can be achieved by discrete variables systems.

\bigskip
\noindent\textbf{2.3 Dense coding in quantum networks}\\
Another important extension is to a quantum dense coding network, where $n>2$ parties share a multipartite entangled state. One strategy is known as distributed quantum dense coding, which generalize standard dense coding to more than one sender and more than one receiver~\cite{1998Bose,2004Brubeta}. For simplicity, we describe the simplest case of a four-party network~\cite{2006Yeo}, where two Senders(S1 and S2) communicate (at most) 4 bits of classical information with two receivers (R1 and R2) by sending in total 2 particles.

When calculating the capacity, distributed dense coding protocols are more complicated for at least three scenarios should be considered: (a) the senders or receivers are distant and communications are not allowed between them; (b) local operations and classical communication (LOCC) are allowed; (c) global operations can be performed by them. These factors do not appear in the standard dense coding scheme where only one sender and one receiver involved. Note that the Holevo bound can be achieved asymptotically for product encodings of the signal states and a complete and orthogonal set for the composite system of all senders can be constructed with a set of local operators of them. So, the capacity of a distributed dense coding cannot be impacted by communications between the senders or their joint operations. For the receivers, however, the above three scenarios result in different capacities, which denoted as $\emph{C}_{lo}$, $\emph{C}_{locc}$, and $\emph{C}_{go}$ respectively. The inequality $\emph{C}_{lo}\leq\emph{C}_{locc}\leq\emph{C}_{go}$ holds, indicating communications between the receivers and their joint operations can boost the rate of communication in a distributed dense coding protocol. Several examples that demonstrate communication~\cite{1989Greenberger} and joint operation~\cite{2006Yeo,2002Verstraete} increasing classical capacity are provided.

Another strategy is controlled dense coding~\cite{2001Hao}, whereby, we describe the simplest case of a three party network. In this scheme, Charlie, rather than acting as a sender or a receiver, controls the dense coding capacity between Alice and Bob. This is made possible by Charlie performing local measurement and broadcasting the result to either Alice or Bob. Sharing the GHZ state $|\Phi_3\rangle=\frac{1}{\sqrt{2}}(|000\rangle+|111\rangle)$ where no entanglement exists between arbitrary two of the three parties, the scheme turns into probabilistic dense coding after Charlie reconstructs entanglement between Alice and Bob. The dense coding capacity is totally controlled by Charlie choosing her measurement basis. With continuous variables systems, the protocol was formulated theoretically~\cite{2002Jing} and demonstrated in lab~\cite{2003cvcontrol} with the help of non-degenerate optical parametric amplifiers. Theoretically, controlled dense coding can be formulated for an arbitrary number of parties~\cite{2009Huang}.

\bigskip
\noindent\textbf{2.4 Dense coding of quantum states}\\
A natural question is if it is possible to assist quantum communication with the help of entanglement, which leads to the topic of dense coding of quantum states~\cite{2004Aram,06Hayden}. In this scenario, Alice aims to communicate an arbitrary 2l-qubit state (known to her) to Bob by transmits only l-qubits with the help of their pre-distributed entanglement. This can never be completed deterministically, or otherwise an arbitrary amount of quantum information could be conveyed by sending a single qubit back and forth. In a probabilistic way, Alice performs a positive-operator valued measurement on her particle and then sends it to Bob, and no Bell state measurement is needed. The method is probabilistic for certain measurement outcome is required to transform the state into the one needed. The success probability is determined by the deviation of the communicated state from the maximally entangled state and can be increased by shared randomness between Alice and Bob. Lower bound on the success probability was derived~\cite{2006Daowen}. Dense coding of quantum states is much related to remote state preparation~\cite{2001Charles,2003}, where Alice remotely prepares Bob's particle in a known state with the help of pre-shared entanglement. A significant difference is that classical communications are prohibited in the former scheme while Alice would broadcast her measurement outcome to Bob in remote state preparation.

\section*{3. Quantum dense coding assisted quantum secure communication}
In the aforementioned dense coding protocols, we focus on increasing the classical capacity at the cost of consuming entanglement resource. Beyond that, quantum resource, as we have known, can assist in realizing secure communication and providing new means of information transmission~\cite{Bennett99,Holevo02,Devetak05,Lloyd97,Barnum98}. We will introduce dense coding based QKD scheme and secure direct communication scheme in this section.

\bigskip
\noindent\textbf{3.1 Dense coding based QKD}\\
Being different from the original dense coding protocol~\cite{1996Bennett}, Alice and Bob perform two eavesdropping checks whenever Alice sends one of the partner particles from an ordered EPR pair sequence to Bob. The first check is used to confirm if there is eavesdropping and the second is to determine if the QKD is successful---if the error rate is below a certain threshold, then the Bell state measurement results are taken as raw keys; otherwise they abandon the results and repeat the procedures from the beginning. The dense coding based QKD protocol~\cite{2002Long,2010Xihan} is secure for a latent eavesdropper Eve can intercept at most one particle from the EPR pairs or her actions are easily detected. For example, Alice and Bob can perceive Eve's strategies based on direct measurement, intercept-resend attack and opaque attack. The protocol exhibits higher efficiency and capacity, compared with the BB84 scheme.

\bigskip
\noindent\textbf{3.2 Dense coding based QSDC}\\
A deterministic cryptographic scheme, also known as QSDC~\cite{2002Beige} can also be developed from dense coding. The important distinction between deterministic and nondeterministic communication is that in the QSDC scheme no classical key is ever established, but rather an inherently quantum-mechanical resource (the shared EPR pairs) takes over the role of the key. The secure direct communication is deterministic for it is possible to communicate the message directly from Alice to Bob. As typical secure direct communication , the ping-pong protocol~\cite{2002KimBostrom,2004QingYu} was proposed. Also, it was also developed from the above QKD protocols based on dense coding, called two-step quantum direct communication protocol~\cite{2003FuGuo}. To realize direct communication, Alice encodes her message directly in the ordered EPR pairs instead or randomly producing them to construct secret key. These protocols, however, are insecure in the case of noisy quantum channel, for a part of message might be leaked to the eavesdropper~\cite{2010Xihan,2003Antoni}. The security problem was further discussed in analogous two-way quantum communication protocols~\cite{03171,03172,03173,03174}, where entanglement is not necessarily needed. Also, high-dimensional secure direct communication was also discussed~\cite{2005Chuan}.

\section*{4. Experimental status and challenges}
In this section, we discuss the major experimental achievements and challenges, which may have not been overcome by now, of dense coding. An experiment of dense coding aims to solve the following problems:
\begin{enumerate}[(1)]
\item Bob can perform a complete Bell state measurement to get the most out of dense coding's potential.
\item A quantum memory is needed to meet the original intention of gaining cheaper "off-peak" rates.
\item The dense coding capacity should exceed the appropriate threshold of $\log_2 d$ bits obtained by sending a d-level state directly.
\end{enumerate}
Some of these items may not be met in the lab. Particularly, a lack of quantum memory may lead to the requirement of two photons arrive at Bob's measurement setup simultaneously in experiment, which in fact despoils quantum dense coding. Another challenge may lie in the implementation of a complete Bell state measurement, a failure of which will lead to an upper bound of the capacity according to the number of Bell state categories that can be identified. Approaching to commercial quantum communication, extra factors have to be taken into consideration:
\begin{enumerate}[(1)]
\item Quantum channels may not be perfect noise free in practical scenarios.
\item It's significant to realize long distance quantum dense coding.
\item Mass data transformed through dense coding might be demanded.
\item The detection efficiency should be high enough to achieve a genuine capacity enhance in practice.
\end{enumerate}

\bigskip
\noindent\textbf{4.1 Photonic qubits}\\
\begin{figure}[tbph]
	\includegraphics [width=0.95\columnwidth]{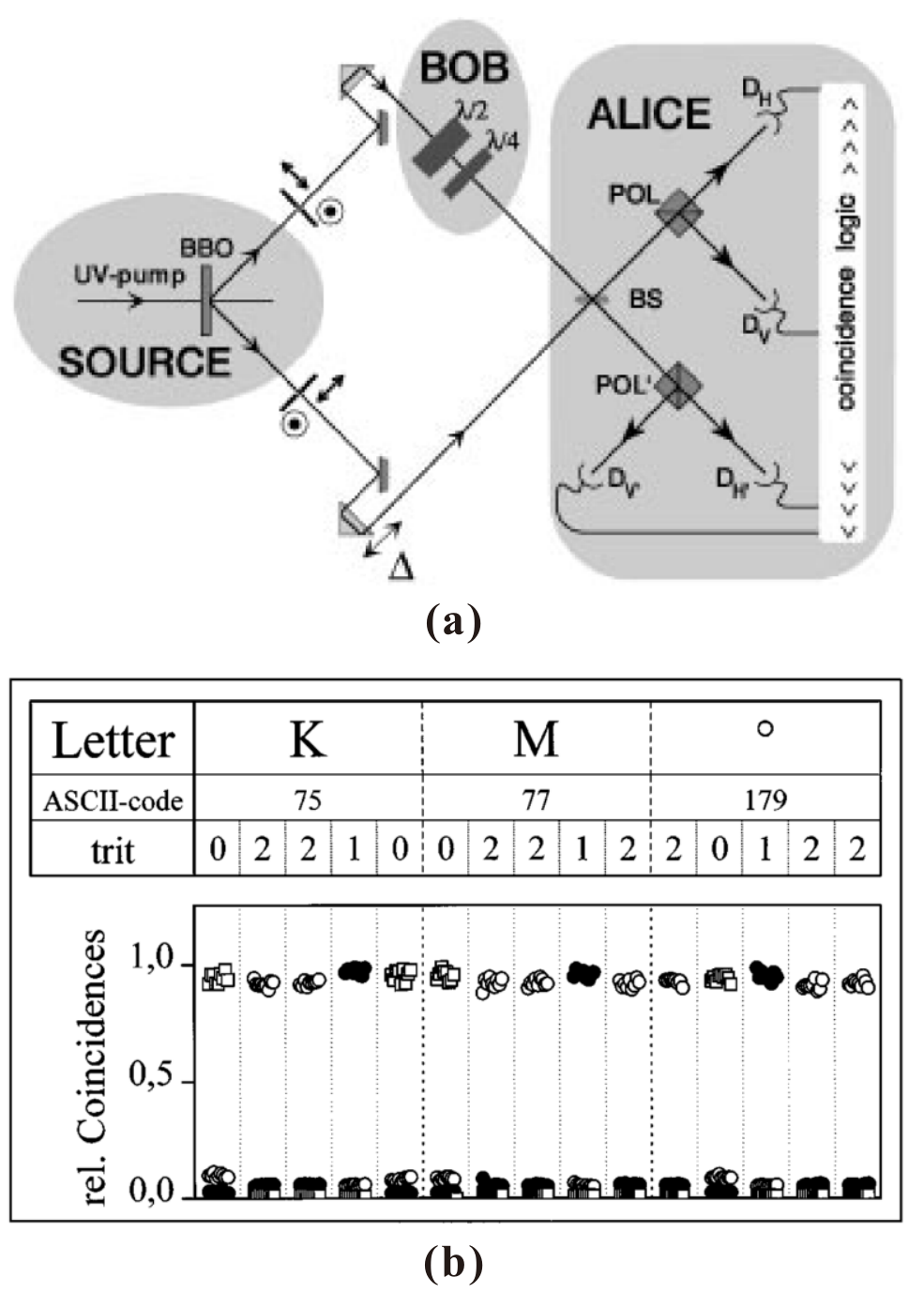}
	\caption{(a) Apparatus of Innsbruck experiment~\cite{96innsbruck}. Polarization entanglement was generated from SPDC and was used to encode the message. Bob could identify 2 out of 4 two-qubit Bell states, while one of the remaining states was chosen as the third signal state.
(b) The ASCII codes for the letters $``{\rm KM}^{\circ}"$ were encoded in 15 trits instead of the 24 bits usually needed.
}
	\label{fig:96exp}
\end{figure}
The very first proof-of-principle demonstration of quantum dense coding is the Innsbruck experiment~\cite{96innsbruck}, where polarization entangled photon qubits from spontaneous parametric down-conversion process (SPDC)~\cite{1995pgkwiat} were used as the information carriers. The information encoding process was implemented by high quality unitary operations with optical polarized elements (half- and quarter- wave plates exactly).  Two photon HOM interferometer~\cite{hom1,hom2} and subsequent polarization analysis were used to distinguish three out of four different messages according to different detection events of the Si-avalanche diodes--state $\Psi^{\pm}$ were registered deterministically by coincidence between different detectors, while state $\Phi^{-}$ (or $\Phi^{+}$ in principle) was signaled with $50\%$ likelihood by coincidence between two detector after an additional beam splitter (Fig.~\ref{fig:96exp}). Photon number resolving detectors are helpful to identify state $\Phi^{-}$ with certainty and made great progress recently~\cite{2017Zolotov,2018Maria}. This experiment obtained an average probability of $92.0\%$ for a correct identification of these signal states and achieved an actual dense coding capacity of 1.13 bit, which showed an apparent signal of exceeding 1 bit by sending a qubit directly. Although far from saying a perfect dense coding trial, this experiment was a great success, since then, great efforts have been made to promote it.

\begin{figure}[tbph]
	\includegraphics [width=0.95\columnwidth]{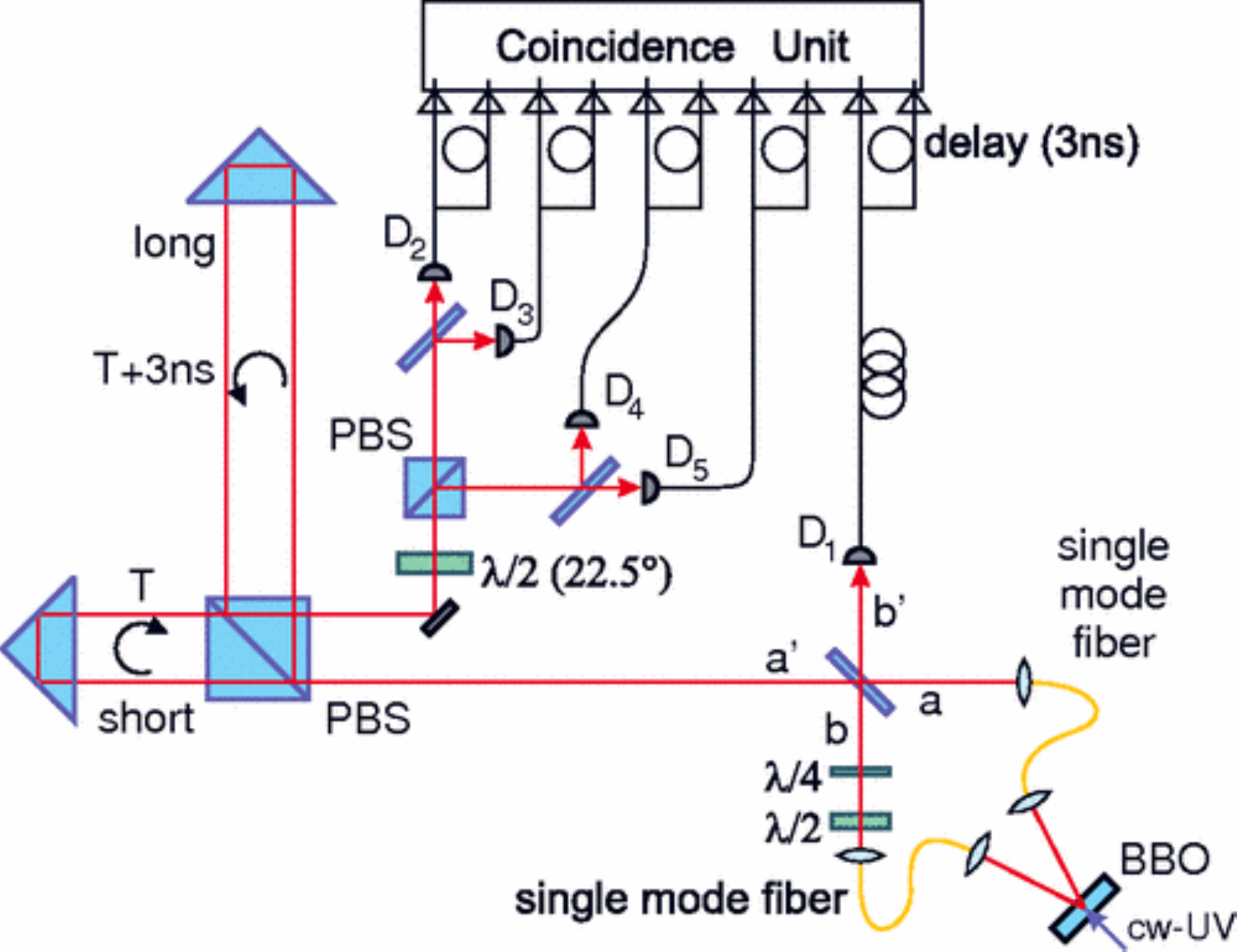}
	\caption{Apparatus of M\"{u}nchen experiment~\cite{06experiment}. Time-energy entanglement was used to realize complete qubit Bell state measurement.}
	\label{fig:06exp}
\end{figure}
\begin{figure}[tbph]
	\includegraphics [width=0.95\columnwidth]{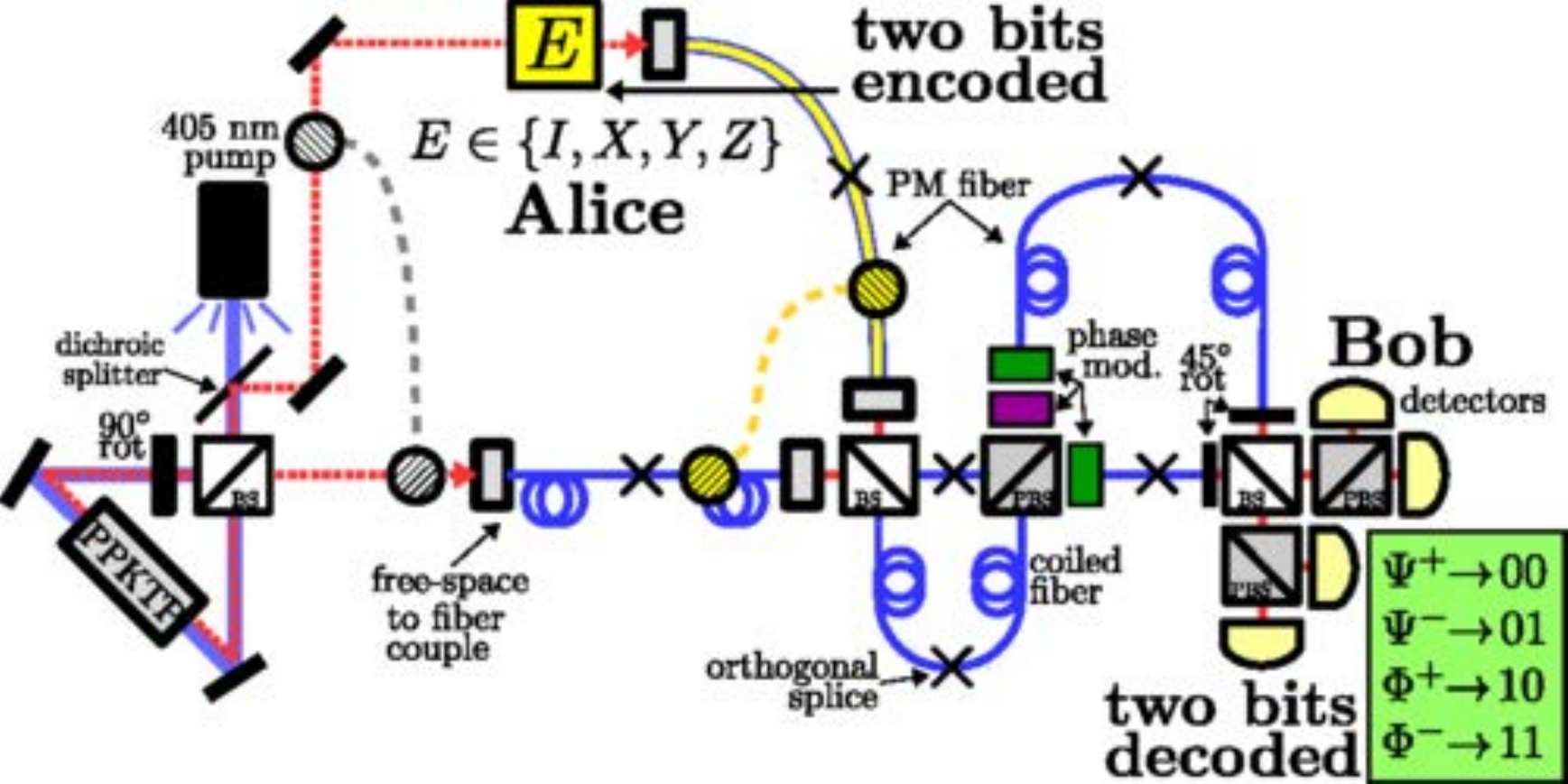}
	\caption{Apparatus of Oak Ridge experiment~\cite{17experiment}. Four temporal modes were used to identify four Bell states without the requirement of PNRDs}
	\label{fig:17exp}
\end{figure}

A significant step to gain higher dense coding capacity is the need to perform a complete Bell state measurement, which is not possible using only linear optics and entanglement in a single degree of freedom~\cite{1999Lutkenhaus,1999Vaidman}. The efficiency of this method is $50\%$--can definitely identify up to 2 of the four Bell states. This reduces the attainable capacity from 2 to $\log_2{3}$ bits when the two identified states and one of the remaining states are used as the signal states. However, entanglement in an extra degree of freedom, known as hyperentanglement~\cite{1997hyper1,2005hyper2}, provides a solution to complete Bell state measurement~\cite{1998hyperbsm,2007hyperbsm}. The method, known as K-W scheme, embeds the Bell states to be identified in a larger Hilbert space by employing extra degrees of freedom Entanglement in these extra degree of freedom are required to permit a second interferometric measurement, which can distinguish between the remaining two Bell states. Using time-polarization hyperentanglement, the M\"{u}nchen experiment~\cite{06experiment}, for the first time, demonstrated photonic dense coding with four signal states based on KW scheme. After the canonical form of Bell state measurement at a beam splitter, an asymmetric Mach-Zehnder interferometer was introduced to distinguish the states $\Phi^{\pm}$ (Fig.~\ref{fig:06exp}). The time-energy correlation resulting from uncertain creation time of the photon pair leaded to different coincidence events of $\Phi^{\pm}$ and thus enabled the distinction of them. The signal states were detected with an average successful probability of $84.2\%$ and the dense coding capacity was $1.18(0.03)$ bits.

\begin{figure}[tbph]
	\includegraphics [width=0.95\columnwidth]{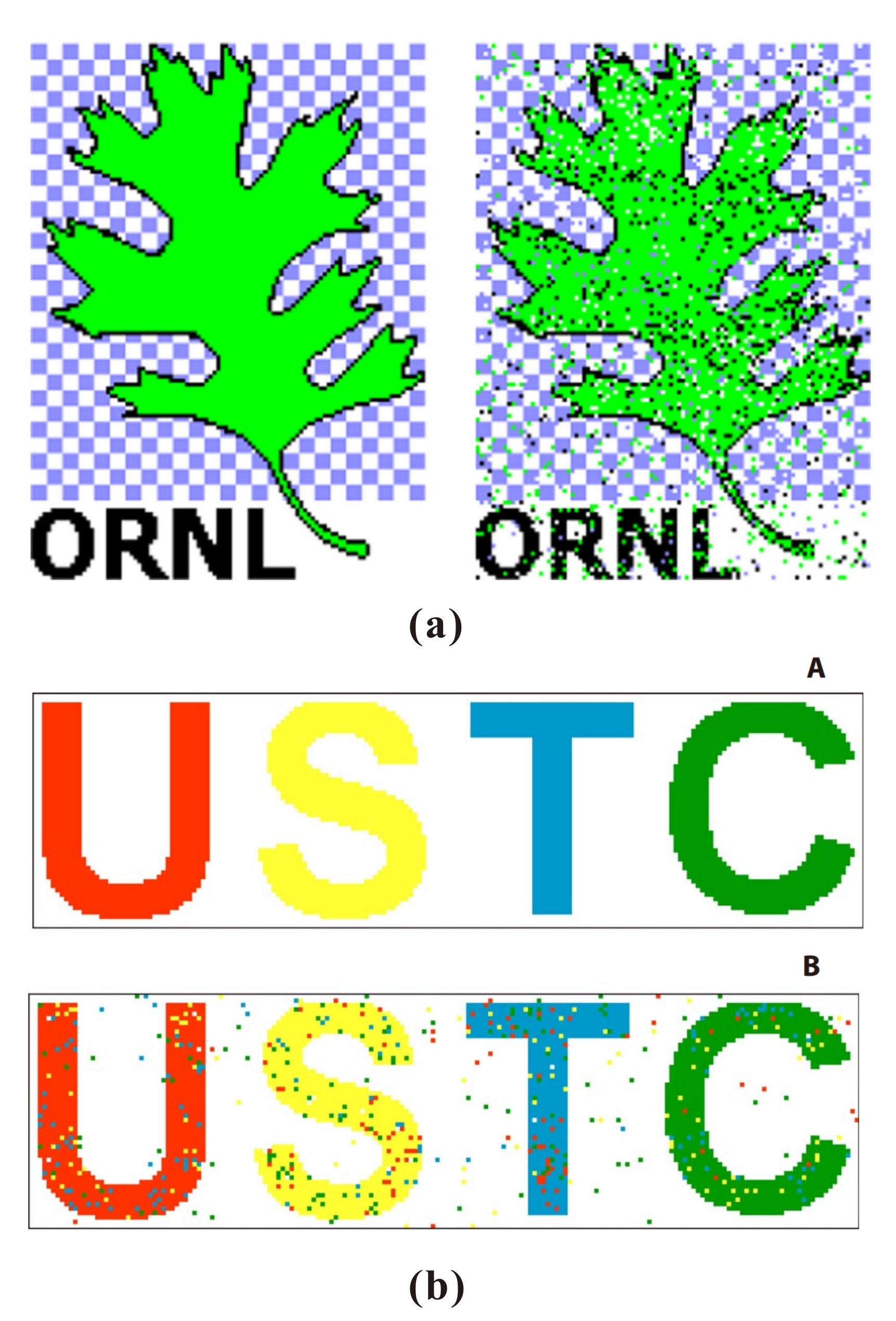}
	\caption{(a) Four color image transmission effect of Oak Rideg experiment~\cite{17experiment} and (b) Five-color image transmission effect of Hefei experiment~\cite{2018experiment}.}
	\label{fig:figure}
\end{figure}

It's also pointed out that photon number resolving detectors were still needed for two photons might be absorbed by a same detector for state $\Phi^{+}$ (with $50\%$ likelihood) and $\Psi^{-}$ (with $25\%$ likelihood). This demand was moved out utilizing a spin-orbit hyperentangled photon pair where orbital angular momentum degree of freedom was introduced to expand the measurement space for breaking the degeneracy of polarization Bell states~\cite{08experiment} or, in turn, polarization are used as an auxiliary to implement complete orbital angular momentum Bell measurement~\cite{2017nankai}. Different complexities to realize these measurements made a big difference between the achieved results--the Illinois experiment achieved an average successful probability of $94.8\%$ and a capacity of $1,630(6)$ bits thus beating a fundamental limit of $\log_2{3}$ bits on the dense coding capacity using only linear optics for the first time, while the Tianjin experiment observed an average success probability of $\sim82\%$ and a capacity of $1.10(4)$ bits which is no better than the Innsbruck experiment~\cite{96innsbruck}. Time-polarization hyperentanglement based dense coding was further developed into the one using four temporal modes in the Oak Ridge experiment~\cite{17experiment} (Fig.~\ref{fig:17exp}). This experiment achieved an average success probability of $95.3\%$ and a capacity of $1.665(0.018)$ bits and, more importantly, paved the way for practical application of quantum dense coding for at least two reasons; firstly, its demonstration over optical fiber links can be easily extended to long distance communication assisted by quantum dense coding, secondly, it showed potential in hybrid quantum-classical transfer protocols by transmitting a $3.4$ kB image with 0.87 fidelity~\ref{fig:figure}(a).

Other methods to achieve more than $50\%$-efficient Bell measurement with linear optical elements include introduction of ancillary entangled~\cite{2011Grice} or unentangled~\cite{2014Fabian} photons and utilization of active optical elements such as squeezers~\cite{2013Hussain}. Using nonlinear interactions, complete Bell measurement has also been demonstrated~\cite{2001Yoon}. These methods, however, may not be as attractive as the the one based on hyperentanglement for quantum dense coding. They may require infinite ancillary photons which could be used to communicate directly or can reach limited Bell efficiency still. Nonlinear interaction solution suffers from poor efficiency.

\begin{figure}[tbph]
	\includegraphics [width=0.95\columnwidth]{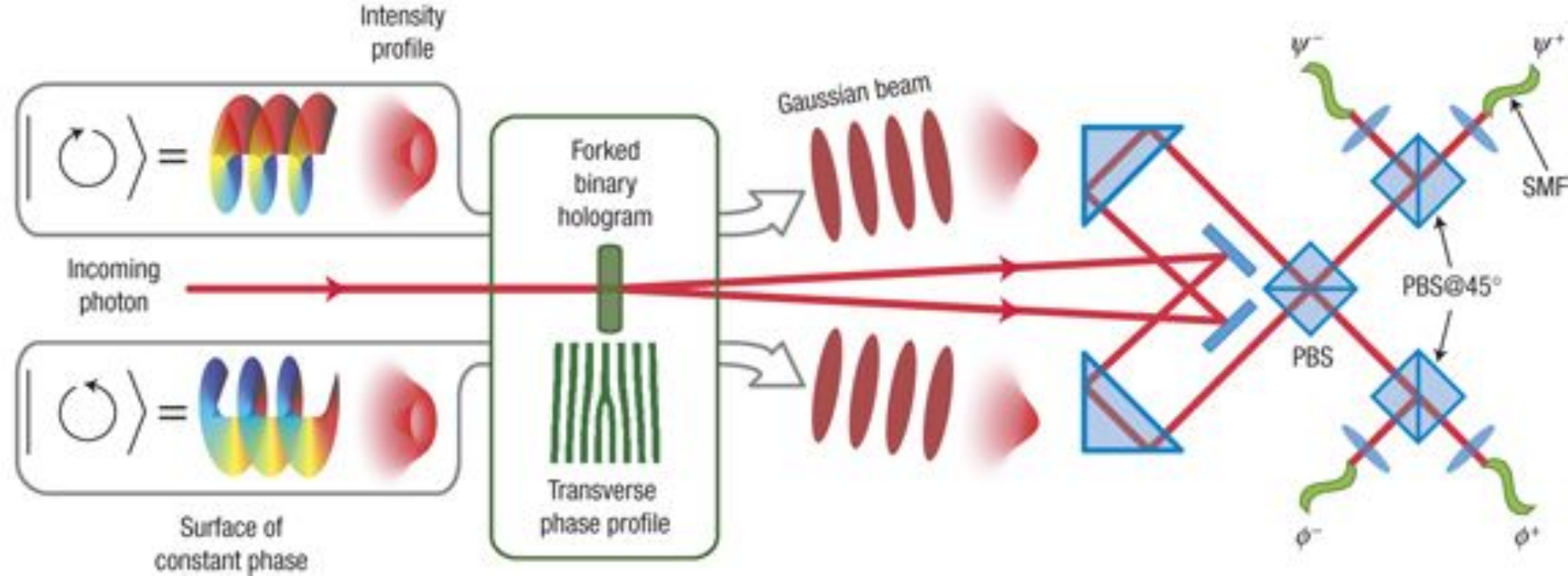}
	\caption{Spin-orbit Bell-state analyser of Illinois experiment~\cite{08experiment}. Complete two-qubit Bell state measurement was achieved by two local spin-orbit Bell-state analysers as shown here.}
	\label{fig:08exp}
\end{figure}

Quantum memory has made great progress~\cite{2010Morgan,2008Appel,2011Specht,2015Manjin,2014Harty}, but has not been used in a quantum dense coding protocol. Optical trombones were used as substitutes for quantum memory in almost all experiments on photonics dense coding. Such an approach played important roles in the proof-of-principle experimental demonstration, but may be not suitable for practical quantum communication where a quantum memory is greatly needed. The demand of quantum memory in dense coding, impressively, can be alternatively simplified to a demand of classical memory~\cite{08experiment}. In the Illinois hyperentanglement experiment, Bob's Bell state measurement was implemented with the help of single photon's spin-orbit Bell states, instead of using two-photon interferometry (Fig.~\ref{fig:08exp}). In this case, Bob can measured his distributed particle locally and store the outcome in a classical memory until Alice sends her particle. This scheme, however, may do harm to communication security, for potential eavesdropper can steal part of information through the intercept-resend attack~\cite{2005intercept}.

\begin{figure}[tbph]
	\includegraphics [width=0.95\columnwidth]{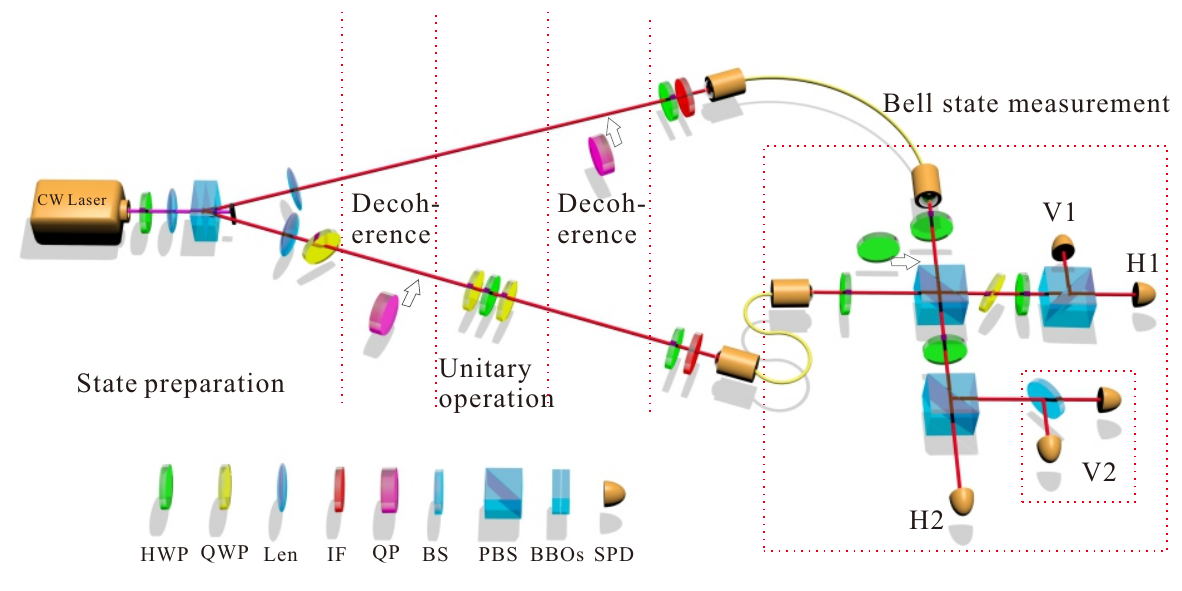}
	\caption{Apparatus of quantum dense coding through non-Markovian noisy channel~\cite{16liu}. Polarization, acting as the system, was coupled to its frequency which acted as the environment. The noisy channel was realized with quatrz plates.}
	\label{fig:16exp}
\end{figure}

To investigate the performance of dense coding through noisy channels, dephasing noise where nonlocal memory effects~\cite{2012Laine} were exploited, was induced~\cite{16liu}. The polarization degree of freedom, acting as an open system, was coupled to its frequency, acting as the non-Markovian~\cite{2008Wolf} environment, to created initial correlations between them (Fig.~\ref{fig:16exp}). Such correlations enabled a nearly perfect capacity ($1.52(0.02)$ bits exactly) with consuming an arbitrary small amount of entanglement (0.163(0.007) ebits). Using a depolarizing channel, a maximal channle capacity of $1.655(0.014)$ was also obtained in a relaxed dense coding scheme where at least two measurement processes were allowed to achieve complete Bell state analysis (two Bell states were identified in a measurement process and the remaining two were done in another)~\cite{2013experiment}.

\bigskip
\noindent\textbf{4.2 Photonic qudits}\\
Compared with qubit entanglement, high-dimensional entanglement~\cite{2015high1,2016high2,2018high3} can benefit quantum communication. For quantum dense coding, $2\log_2{d}$ bits of information can be transmitted by sending a single qudit particle. The aforementioned hyperentanglement assisted dense coding may be compared with the one based on high-dimensional entanglement, for hyperentanglement itself can be treated as high-dimensional entanglement after recoding~\cite{2018genuine}. A significant difference exists in where the information is encoded: message is encoded on a qubit subspace of a certain degree of freedom for the hyperentanglement scheme while it is encoded on the whole space for the high-dimensional scheme.

\begin{figure}[tbph]
	\includegraphics [width=0.95\columnwidth]{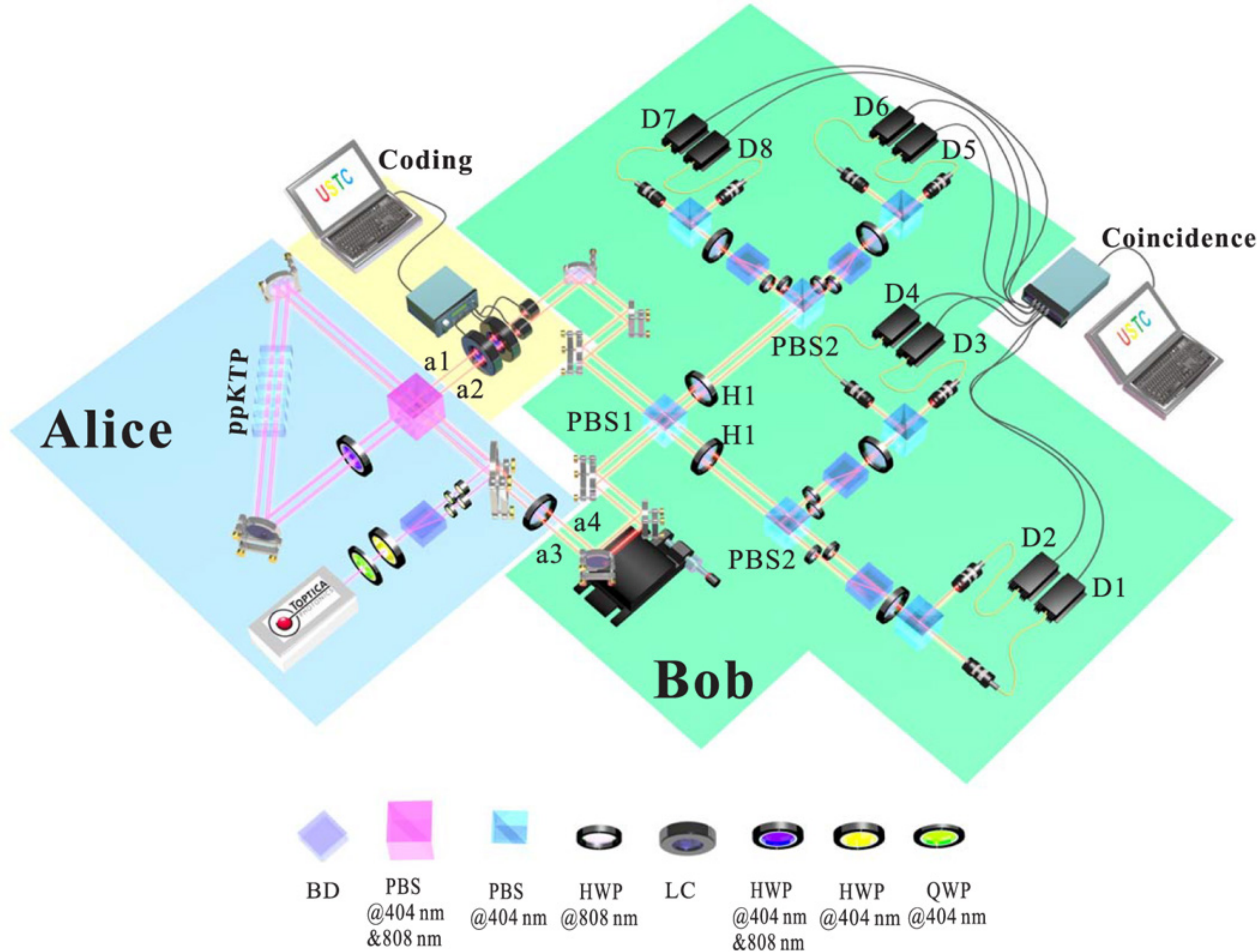}
	\caption{Apparatus of Hefei experiment~\cite{2018experiment}. Polarization and path were used to encode the 4-dimensional entangled states. Bob's setup could be used to classify the 16 4-dimensional Bell states into 7 classes, each class containing at least 2 states.}
	\label{fig:18exp}
\end{figure}

The first high-dimensional dense coding demonstration is the Hefei experiment~\cite{2018experiment}, where photonic polarization and path were used to encode 4-level entanglement after a SPDC. For a 4-level system, there are 16 Bell states which can be used as the signal states to encode the transmitted message. After a entangled ququart photon pair was distributed, Alice performs local ququart operations with the help of several computer-controlled liquid crystal variable retarders to realize state transformations between 5 selected signal states. Bob can then retrieve the message by performing four dimensional Bell state measurement on his distributed photon and Alice's photon (Fig.~\ref{fig:18exp}).

A major innovation of the experiment is it provided a partial answer to the realization of high-dimensional Bell state measurement, while a complete 4-level Bell state measurement is impossible within linear optics. To achieve this, another HOM-like interferometry was used after the standard one for 2 dimensional Bell state measurement, which realized the separation the 16 Bell states into 7 classes (each contains at least 2 states). The present setup enabled encoding message with 6 signal states and the remain one can be utilized as long as PNRDs are used. Analogous to K-W scheme which deals with the 2-dimensional case, complete high-dimensional Bell state measurement can be reached with hyperentanglement or ancillary particle assistance~\cite{2018haozhang}.

The Hefei experiment achieved an average success probability of $93.5\%$ and a capacity of $2.09\pm0.01$ bits, which exceeds the fundamental limit of dense coding with qubit system and also the limit of transmitting a single ququart directly. A five-color image was also transferred at a rate of 0.5 Hz with $0.95$ fidelity~\ref{fig:figure}(b). The high capacity benefitted not only from more signal states but also from the choice of polarization and path to prepare high-dimensional entanglement, for it can be prepared, operated, and measured with near unit fidelity.

\bigskip
\noindent\textbf{4.3 Continuous variables systems}\\
Although entanglement resoures with high detection efficiency have been reported~\cite{2015highdetection,2018highdetection}, previous photonic dense coding experiments didn't consider the detection efficiency when calculating the capacities. The actual capacities of the experiments thus can hardly exceed the fundamental bound of 1 bit by directly transmitting a single qubit. The first and simplest resolution of inefficiency came from the use of continuous variables (CV) systems, for nearly $100\%$ detection efficiency can be achieved here. Moreover, the CV version of complete Bell state measurement can easily be implemented with two methods as mentioned above.

\begin{figure}[tbph]
	\includegraphics [width=0.95\columnwidth]{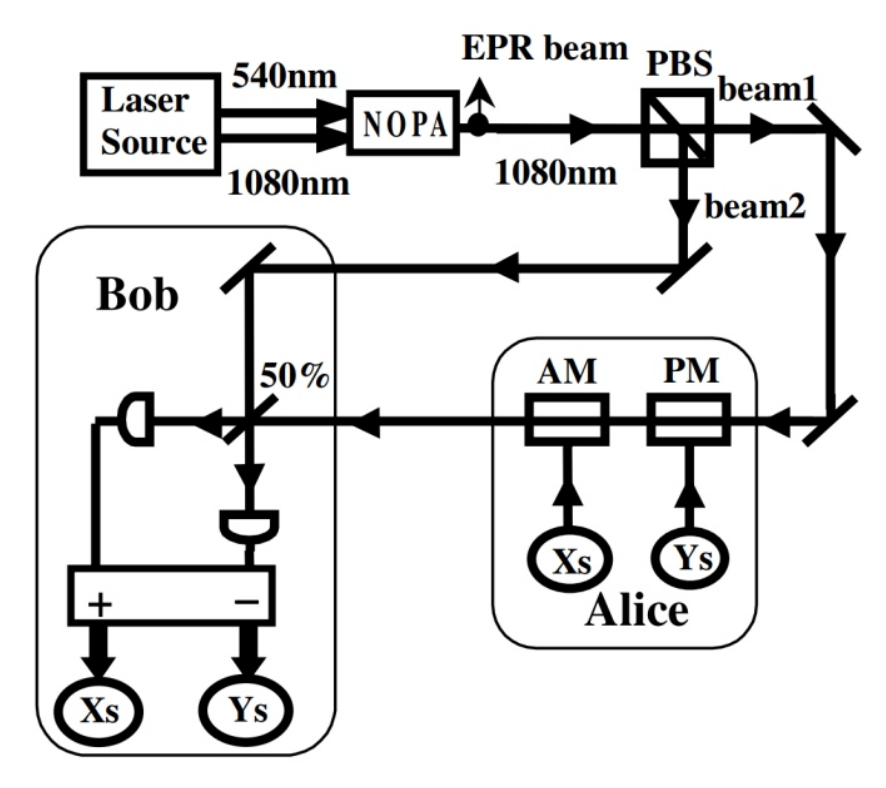}
	\caption{Apparatus of Taiyuan experiment~\cite{2002cvexperiment}. Two-mode squeezed light from NOPA was used to realize continuous variables dense coding. Bob decoded the message by performing a direct measurement of the Bell states.}
	\label{fig:02cv}
\end{figure}

To accomplish a CV quantum dense coding, EPR beam, two-mode squeezed light here, of approximately $70\mu W$ with correlated amplitude quadratures and anticorrelated phase quadratures produced by a continuous nondegenerate optical parametric amplifier was used in the Taiyuan experiment~\cite{2002cvexperiment}. The EPR beams were shared by Alice and Bob to accomplish the protocol. Two bits of classical information were encoded on the amplitude and phase quadratures, which were done by implementing a phase-space offset on the half of EPR beams at Alice. Bob decoded the message by performing a direct measurement of the Bell states~\cite{2000Zhang} on the whole EPR beams--imposing a $\pi/2$ phase between the beams and then combining them on a $50\%$ beamsplitter, the sum and difference of the photocurrent at the two outputs of the beamsplitter were recorded as the retrieved messages (Fig.~\ref{fig:02cv}). Although with imperfect EPR beams, the encoded amplitude and phase messages were retrieved with an average signal-to-noise ratios of $3.8dB$ beyond that of the shot noise limit.

\begin{figure}[tbph]
	\includegraphics [width=0.95\columnwidth]{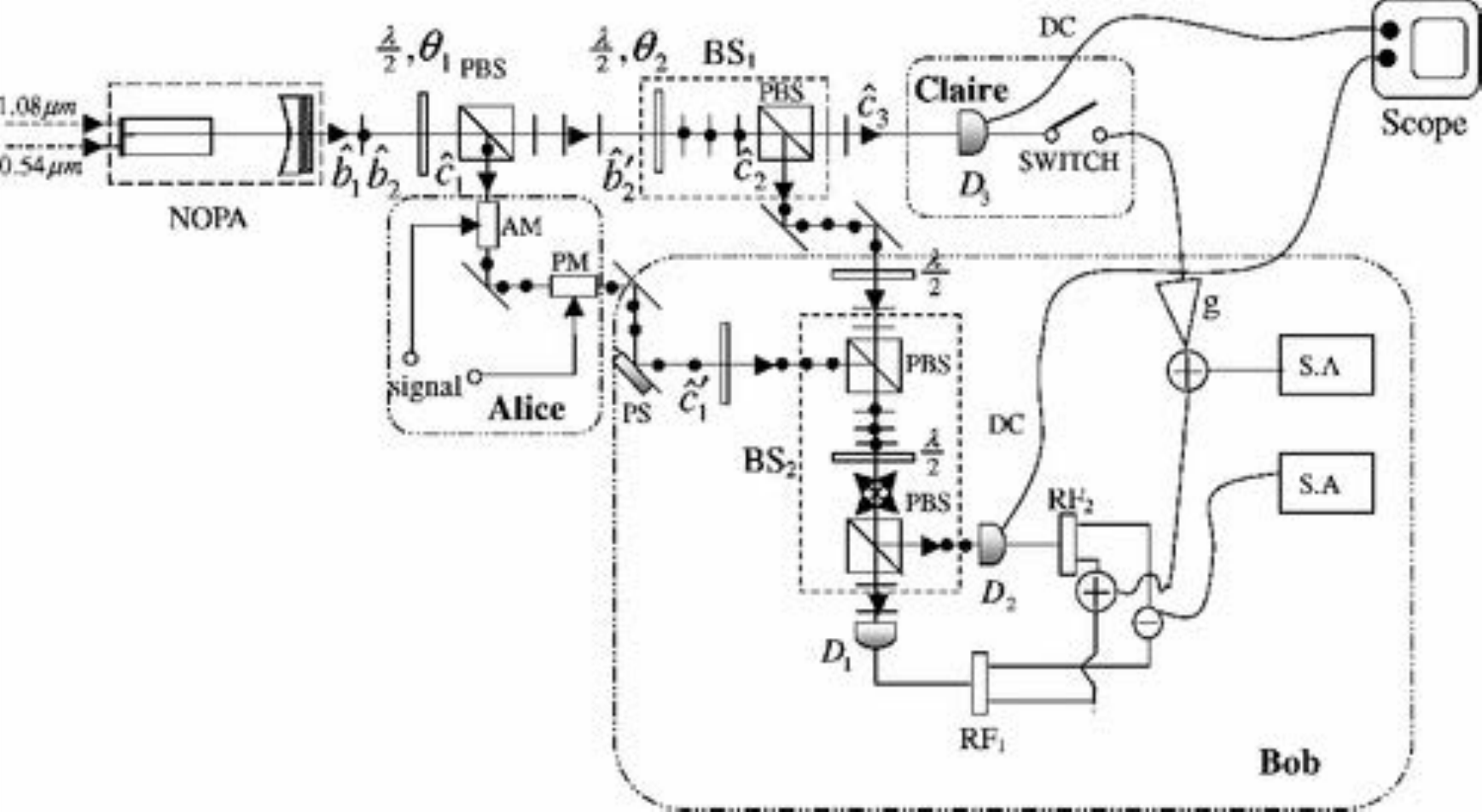}
	\caption{Apparatus of controlled dense coding with tripartite entanglement on a cv system~\cite{2003cvcontrol}. Charlie, as a controller, could control the channel capacity between Alice and Bob.}
	\label{fig:03cv}
\end{figure}

The experiment was later developed into a controlled dense coding for continuous variables, where tripartite entanglement was distributed between Alice, Bob and Charlie~\cite{2003cvcontrol}. In such a scheme, Charlie acts as a controller who can adjust the capacity between Alice and Bob if they are communicating through a dense coding channel. By distributing a two-mode squeezed state to three parties, the state resulted in a three-mode position eigenstate with the quantum correlations of total position quadratures and relative momentum quadratures~\cite{2002cvtri}. To control Alice and Bob's communication, Charlie detected his distributed mode and sent the result to Bob, who can then modify his own Bell measurement results to achieve higher capacity (Fig.~\ref{fig:03cv}).

\begin{figure}[tbph]
	\includegraphics [width=0.95\columnwidth]{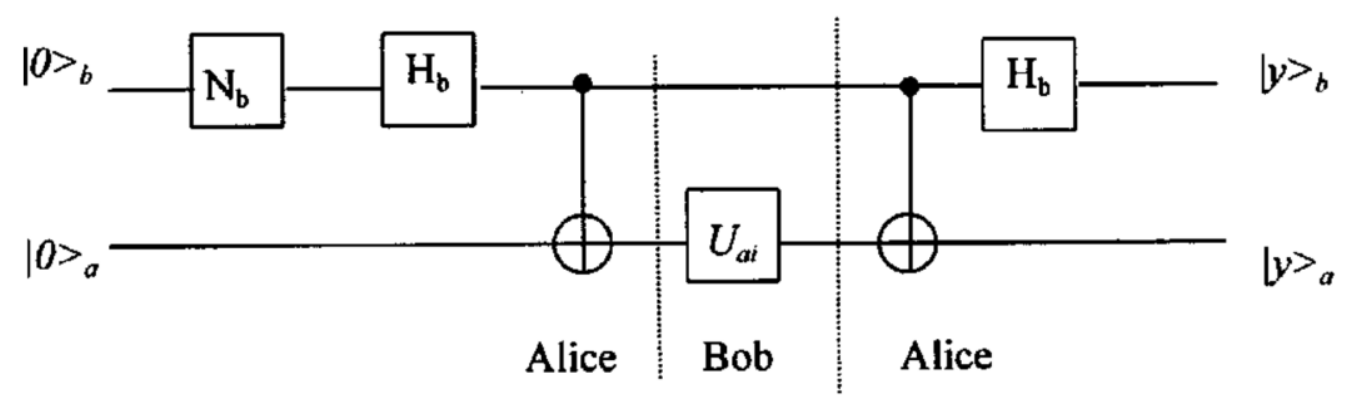}
	\caption{The quantum circuit for dense coding with NMR~\cite{2000nmr}. H and N denote the Walsh-Hadamard gate and NOT gate.}
	\label{fig:00nmr}
\end{figure}

\bigskip
\noindent\textbf{4.4 Nuclear magnetic resonance}\\
Dense coding was also demonstrated in NMR~\cite{2000nmr}, where the qubits are nuclear spins. The two-spin system was chosen to be composed of a hydrogen nucleus (H) and a carbon nuclei (C) in the molecule of carbon-13 labeled chloroform. Employing an spatial averaging method, the qubits C and H were prepared to pure state $|00\rangle$, before entanglement between them were created through spin-spin interactions. After the encoding process of Alice, Bob performed a complete Bell state measurement by the action of a controlled Not gate and a following Hadamard gate. The gate series enabled different Bell states to be projected on different computational basis which could be easily distinguished by projective measurement (Fig.~\ref{fig:00nmr}). Multi-parties quantum dense coding was also realized in a sample of carbon-13 labeled alanine, where 2 senders transmitted totally 3 bits of information to the receiver by sending only 2 qubits~\cite{2004nmr}. A major problem of NMR dense coding is that communication only takes place between spins in {\AA} distance as the nuclear spins in a molecule are coupled through chemical bond.

\bigskip
\noindent\textbf{4.5 Atomic system}\\
Dense coding and also QSDC based on dense coding may be performed within atomic quantum systems, including trapped atomic qubits~\cite{2004atom} and atomic ensembles~\cite{2017weizhang}. Atomic systems are good candidates for practical quantum memory~\cite{2011Specht,2014Harty}, so they would play an important role in future quantum communication network, including quantum dense coding of course.

\begin{figure}[tbph]
	\includegraphics [width=0.95\columnwidth]{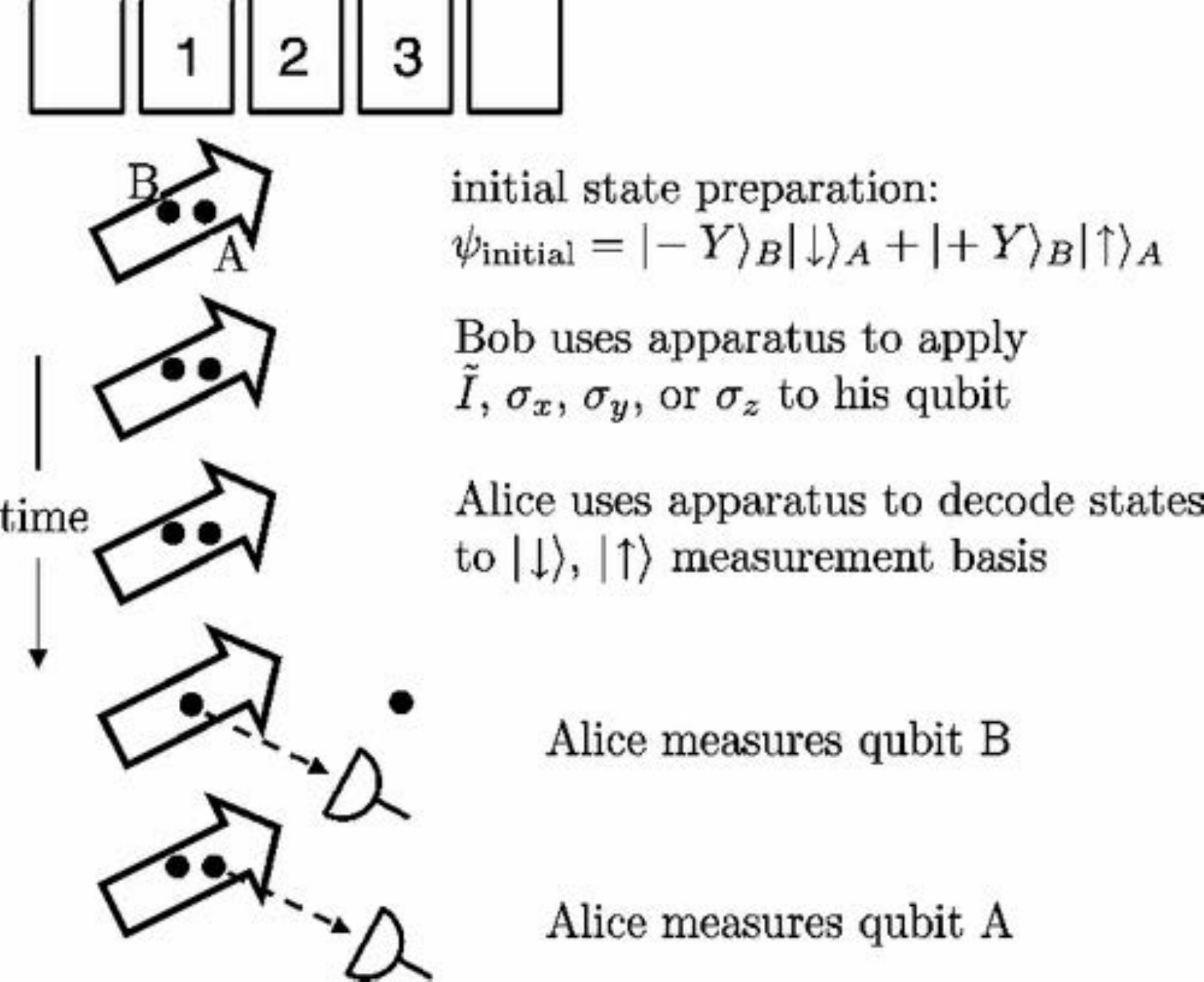}
	\caption{Schematic diagram of Colorado experiment~\cite{2004atom}. Bell state measurement was performed by measuring qubit B and A sequentially--qubit A was transferred to an auxiliary zone while qubit B was measured; qubit B was pumped to dark state when qubit A was measured.}
	\label{fig:04atom}
\end{figure}

The Colorado experiment~\cite{2004atom} considered two beryllium ions ($^{9}Be^{+}$) confined in a multizone linear re-Paul trap, where the ground-state hyperfine levels of $^{9}Be^{+}$ provided with the qubits. The ions were coupled through short-range Coulomb interacitons, so the trapped ions dense coding procedure was much analogous to a NMR one, where Alice encoded her message and turned over the apparatus to Bob, who then decoded the information using both qubits. Single qubit gates and a two-qubit universal logic gate were used to construct the whole circuite to implement a trapped ions dense coding. A single-qubit gate was accomplished with stimulated Raman transitions excited with two laser beams, while the two-qubit gate was configured by applying state dependent optical dipole forces. Complete Bell state measurement on two ions can be accomplished by firstly projecting the four states on different two-qubit computational basis and then applying resonance fluorescence on two qubit independently (Fig.~\ref{fig:04atom}). As a result, an average success probability of $85.0\%$ and a dense coding capacity of 1.16(1) bits were observed in this experiment. High fidelity single-qubit gates ($>0.9999$) and two-qubit gates ($>0.999$) have been reported~\cite{2016gate1,2016gate2}, which would enable trapped atomic dense coding with a nearly perfect capacity.

\begin{figure}[tbph]
	\includegraphics [width=0.95\columnwidth]{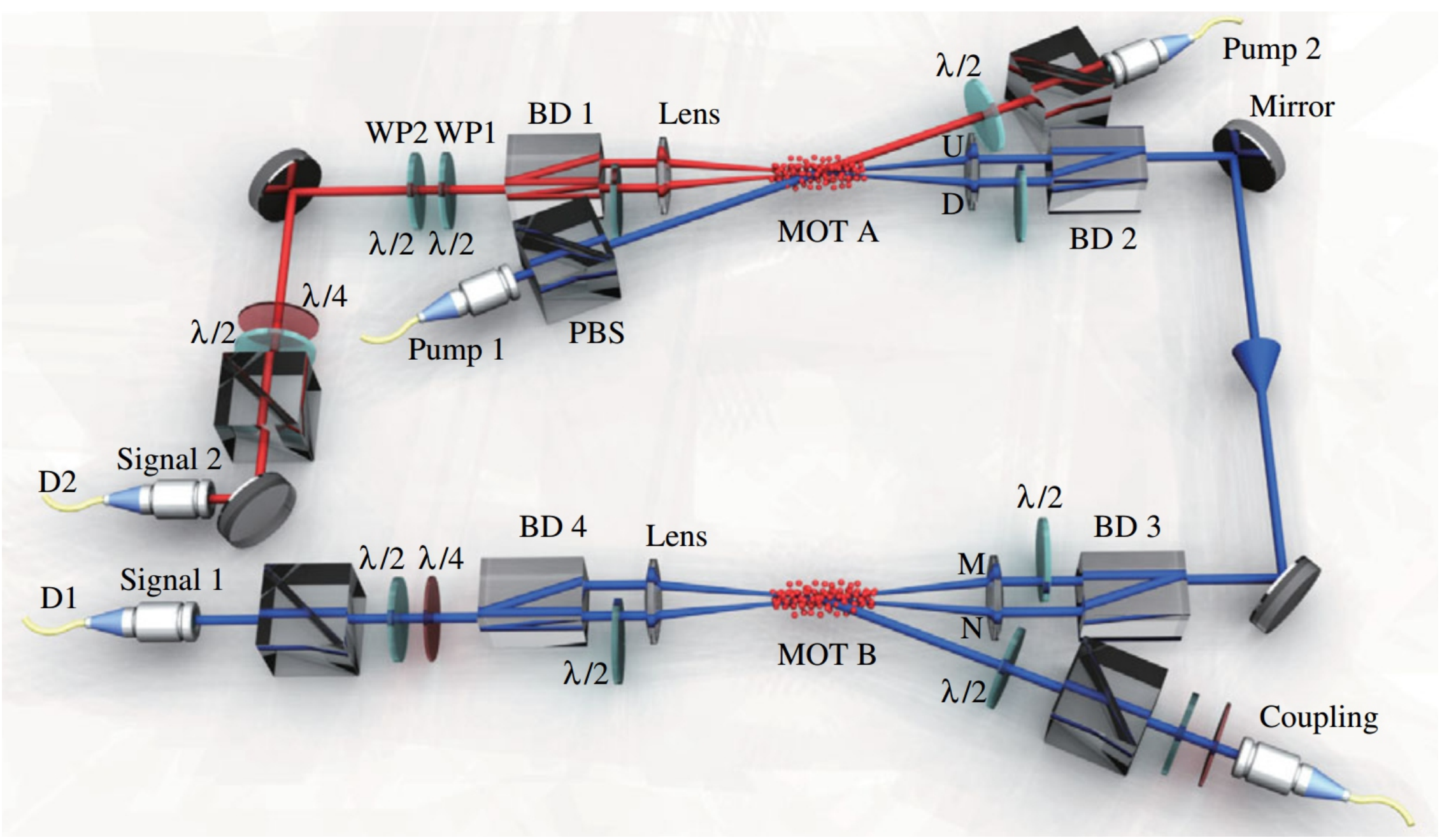}
	\caption{ Apparatus of QSDC~\cite{2017weizhang}. Rubidium atomes in MOT A and MOT, acting as memories, were entangled and their retrieved photons were measured to decode the message. The storage times for memories in MOT A and MOT B were $\bigtriangleup t_1=50$ ns and $\bigtriangleup t_1=120$ ns.}
	\label{fig:16qsdc}
\end{figure}

A QSDC protocol~\cite{2003FuGuo} can be viewed as quantum dense coding with two extra channel security checks and the whole process is performed with N pairs of pre-distributed entangled states. QSDC has been demonstrated between two ensembles of rubidium ($^{85}Rb$) atomes trapped in magneto-optical traps (MOT)~\cite{2017weizhang}. The atom-photon entanglement was generated by sending a pump pulse through the first ensemble and the photon was then delivered to the second ensemble for storage, thus establishing the pre-distributed memory-memory entanglement for Alice and Bob. Alice encoded her message on the retrieved photon from the first ensemble and sent it to Bob, who then decoded the message by reconstructing the density matrix of the photon and the other retrieved photon from the second ensemble (Fig.~\ref{fig:16qsdc}). The storage time for the two atomic ensembles were 50 ns and 120 ns. The experiment achieved an average message-retrieving fidelity of $90.1\%$, which was defined as the average fidelity of the retrieved states.
\begin{table*}[btph]
\caption{Reported experimentally achievements of quantum dense coding. $N_s$: number of signal states; $P_s$: average detection success probability; ODL: optical delay line.}
\begin{tabular}{c|c|c|c|c}
  \toprule
  \hline
  Quantum technology & \quad $N_s$ \qquad & \quad $P_s$ ($\%$) \qquad & \quad capacity (bits) \qquad & \quad memory \quad\\
  \hline
  Polarization qubits$^{15}$ & 3 & $92.0$ & $1.13$ & ODL \\
  \hline
  \quad Time-bin assisted polarization$^{20}$ \quad & 4 & $82.4$ & $1.18$ & ODL  \\
  \hline
  OAM assisted polarization$^{21}$ & 4 & $94.8$ & $1.63$ & ODL  \\
  \hline
  Polarization assisted OAM$^{22}$ & 4 & $82.0$ & $1.10$ & ODL  \\
  \hline
  \quad Time-bin assisted polarization$^{23}$ \quad & 4 & $95.3$ & $1.665$ & ODL  \\
  \hline
  Qubits in noisy channel$^{82}$ & 3 & $-$ & $1.52$ & ODL  \\
  \hline
   Photonic qudits$^{24}$ & 5 & $93.5$ & $2.09$ & ODL  \\
  \hline
   Optical modes$^{16}$ & 4 & $-$ & $-$ & ODL  \\
  \hline
   NMR$^{17}$ & 4 & $\approx90$ & $-$ & -  \\
  \hline
  Trapped atoms$^{18}$ & 4 & $85.0$ & $1.16$ & $>0.7\mu s$  \\
  \hline
  Atomic ensembles$^{19}$ & 4 & $90.1$ & $-$ & 120~ns  \\
  \hline
  \bottomrule
\end{tabular}
\label{table:capacity}
\end{table*}

\section*{5. Discussion and outlook}


Tab.~\ref{table:capacity} summarizes the reported experimental performances of quantum dense coding. Despite the great advances that have been achieved, as mentioned in the main text, with the use of various technologies (from photonic systems and optical modes to NMR and atomic systems), it seems not possible to construct a practical dense coding apparatus with only one of these substrates, for each of them has its fatal weakness. Photonic qubits enable high detection success probability of Bell state and it is possible develope high-dimensional dense coding protocols based on photon systems. More over, photon, in particular polarization qubit, provides the best carriers for long-distance quantum dense coding over fibres or free-space links. The problem is, photon may suffer high losses associated with spreading. Fortunately, quantum repeater~\cite{repeater} and satellite-to-ground quantum communication~\cite{16satellite} shed light on solution to this field. Continuous variables can achieve high detection efficiency and are robust to noisy channels. However, continuous variables and also photonic systems will ask for help to provide quantum memory from other systems. Atomic systems~\cite{2011Specht,2014Harty} and also solid-state systems~\cite{2010Morgan,2008Appel,2015Manjin}, inversely, have the potential to construct well-rounded quantum memories, but are not useful for long-distance communication. Based on the discussion above, we conclude that future practical quantum dense coding apparatus may depend on integration of photonic and optical network with quantum memories based on atomic and solid-state systems, in line with hybrid approaches to quantum teleportation~\cite{np2015} and quantum information~\cite{nphys2015}.

\section*{ACKNOWLEDGMENTS}
This work was supported by the National Key Research and Development Program of China (Nos. 2017YFA0304100, 2016YFA0301300, 2016YFA0301700), NSFC (Nos. 11774335, 11874345, 11821404), the Key Research Program of Frontier Sciences, CAS (No. QYZDY-SSW-SLH003), the Fundamental Research Funds for the Central Universities, and Anhui Initiative in Quantum Information Technologies (Nos. AHY020100, AHY060300).


\end{document}